\documentstyle[11pt,newpasp,twoside,epsf]{article}
\begin{document}
\title{Numerical simulations of turbulent dynamos}
 \author{Axel Brandenburg}
\affil{NORDITA, Blegdamsvej 17, DK-2100 Copenhagen \O, Denmark; and\\
Department of Mathematics, University of Newcastle upon Tyne, NE1 7RU, UK}

\begin{abstract}
Using a periodic box calculation it is shown that, owing to helicity
conservation, a large scale field can only develop on a resistive
timescale. This behaviour can be reproduced by a mean-field dynamo
with $\alpha$ and $\eta_{\rm t}$ quenchings that are equally strong and
`catastrophic'.
\end{abstract}

\section{Introduction}

After the original papers by Vainshtein \& Cattaneo (1992) and Cattaneo
\& Vainshtein (1991) about `catastrophic' $\alpha$ and $\eta_{\rm t}$
quenching, respectively, there has been a continuous debate about the
existence and relevance of $\alpha$-effect dynamos in astrophysics,
i.e.\ in the high $R_{\rm m}$ regime. Their results were supported by
calculations that did not however have dynamo action (see also Cattaneo
\& Hughes 1996). In the present paper we show, using a periodic box
calculation that, owing to helicity conservation, a large scale field
can only develop on a resistive timescale, and that this behaviour can
be reproduced by a mean-field dynamo with $\alpha$ and $\eta_{\rm t}$
quenchings that are equally strong and `catastrophic' in the sense that
quenching sets in once the large scale field reaches the fraction $R_{\rm
m}^{-1/2}$ of the equipartition value.

\section{Helicity constraint and catastrophic quenching}

Large scale magnetic field generation has long been associated with
helicity (e.g., Moffatt 1978, Krause \& R\"adler 1980). However,
the large scale field generated by turbulence with kinetic helicity
generally possess also magnetic helicity. This is a conserved quantity and
characterizes the linkage of magnetic flux structures with themselves. If
the magnetic helicity is zero initially within a certain volume, it can
only change if there is a loss of magnetic helicity of preferentially
one sign through the boundaries (Blackman \& Field 2000, Kleeorin et al.\
2000). Magnetic helicity can also change resistively, allowing the large
scale magnetic helicity to grow while the small scale magnetic helicity
is being dissipated, but this is a slow process if the magnetic Reynolds
number is large. Using only helicity conservation and the assumptions
that the large scale field, $\overline{\mbox{\boldmath $B$}}$, possesses
helicity, one can show that (Brandenburg 2000, hereafter B2000)
\begin{equation}
\langle\overline{\mbox{\boldmath $B$}}^2\rangle\approx k_{\rm f}\ell_{AB}
B_{\rm eq}^2\left[1- \exp(-2\eta k_1^2 (t-t_{\rm sat})\right].
\label{helconst}
\end{equation}
Here, $t_{\rm sat}=\lambda^{-1}\ln(B_{\rm fin}/B_{\rm ini})$ is the time
when the field at small and intermediate scales saturates, $B_{\rm ini}$
and $B_{\rm fin}$ are initial and final field strengths, $\lambda$ is the
kinematic growth rate of the dynamo, $B_{\rm eq}$ is the equipartition
field strength, and $k_{\rm f}$ is the wavenumber of the forcing.

Equation~(1) is rather general and independent of the actual
model of field amplification. If the field is not fully
helical, then $\ell_{AB}=|\langle\overline{\mbox{\boldmath
$A$}}\cdot\overline{\mbox{\boldmath
$B$}}\rangle|/\langle\overline{\mbox{\boldmath $B$}}^2\rangle$ will
be less than $1/k_1$, where $k_1=2\pi/L$ is the smallest possible
wavenumber. Thus, the final field amplitude will be reduced. In any case
however the time it takes to reach this final amplitude scales with the
resistive time, $(\eta k_1^2)^{-1}$, where $\eta$ is the microscopic
(not the turbulent!) value.

Equation~(1) suggests that, after $t=t_{\rm sat}$ the large scale
field grows at first linearly, $\langle\overline{\mbox{\boldmath
$B$}}^2\rangle=2\eta k_1^2 (t-t_{\rm sat})$, and then slowly
approaches $B_{\rm fin}$. In Fig.~1 we compare the evolution of
$\langle\overline{\mbox{\boldmath $B$}}^2\rangle$ for Run~3 of B2000 with
the result of Eq.~(1). In this run we have $B_{\rm ini}=2\times10^{-9}$,
$B_{\rm fin}=0.342$, and $\lambda=0.067$, so $t_{\rm sat}=283$.

\begin{figure}[h!]\plotone{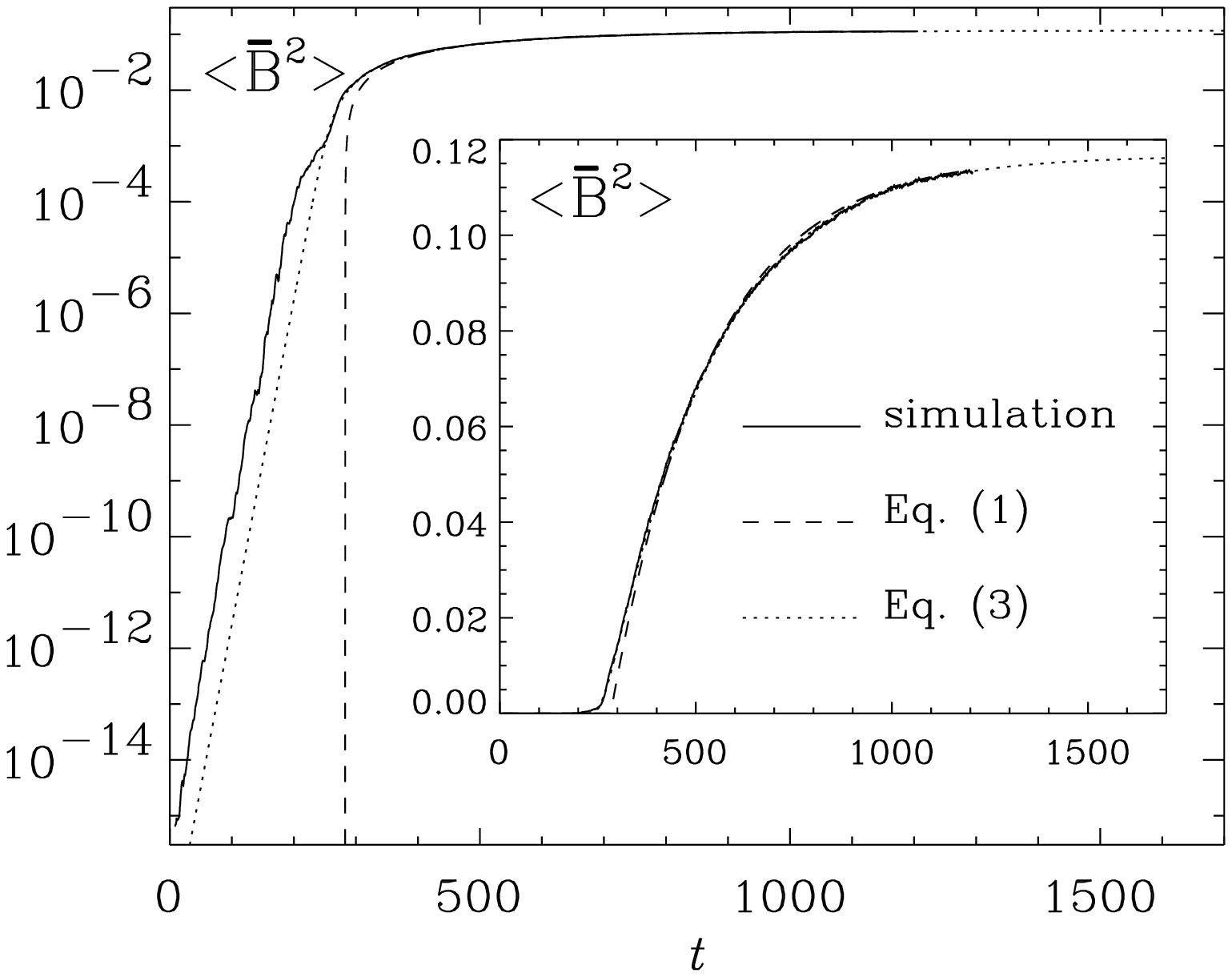}\caption[]{
Evolution of $\langle\overline{\mbox{\boldmath $B$}}^2\rangle$ for Run~3
(solid line) compared with Eq.~(1) (dashed line). The dotted line gives
the result for an $\alpha^2$ mean-field model (discussed below). The
inset shows the evolution on a linear scale. Note the excellent agreement
between simulation and the analytic fits during the saturation phase.
$u_{\rm rms}=0.18$, $k_1=1$, $k_{\rm f}=5$, $\eta=0.002$.
}\label{Fpjbm_decay_nfit}\end{figure}

\section{How to model this?}

Exactly the same resistive asymptotic behaviour can be {\it reproduced}
using an $\alpha^2$-dynamo with simultaneous $\alpha$ and $\eta_{\rm
t}$-quenching of the form
\begin{equation}
\alpha={\alpha_0\over1+\alpha_B\overline{\mbox{\boldmath
$B$}}^2\!/B_{\rm eq}^2},\quad \eta_{\rm t}={\eta_{\rm
t0}\over1+\eta_B\overline{\mbox{\boldmath $B$}}^2\!/B_{\rm eq}^2},
\label{quench_both}
\end{equation}
where $\alpha_B=\eta_B$ is assumed. Assuming that the magnetic
energy density of the mean field, $\overline{\mbox{\boldmath
$B$}}^2$, is approximately uniform (which is well obeyed in the
simulations) we can obtain the solution $\overline{\mbox{\boldmath
$B$}}=\overline{\mbox{\boldmath $B$}}(t)$ of the $\alpha^2$-dynamo
equations in the form (B2000)
\begin{equation}
\overline{\mbox{\boldmath $B$}}^2/(1-\overline{\mbox{\boldmath
$B$}}^2\!/B_{\rm fin}^2)^{1+\lambda/\eta k_1^2} =B_{\rm
ini}^2\,e^{2\lambda t},
\label{approx_quench}
\end{equation}
where $\lambda=|\alpha_0|k_1-\eta_{\rm T0}k_1^2$ is the kinematic growth
rate, and $\eta_{\rm T0}=\eta+\eta_{\rm t0}$. The reason the field
saturates in spite of simultaneous quenching of $\alpha$ and $\eta_{\rm
t}$ is that there remains still the microscopic diffusivity $\eta$,
which is not quenched, preventing therefore indefinite growth. This
determines the saturation field strength in terms of the quenching
parameters $\alpha_B$ and $\eta_B$ and yields
\begin{equation}
\alpha_B=\eta_B={\lambda\over\eta k_1^2}
\left({B_{\rm eq}\over B_{\rm fin}}\right)^2,
\label{quench_formula}
\end{equation}
where $B_{\rm eq}^2=\mu_0\langle\rho\mbox{\boldmath $u$}^2\rangle$
is characteristic of the magnetic energy at small scales. To a good
approximation the spectral energy in the small scales is close to that
in the large scales, so $k_1 B_{\rm fin}^2\approx k_{\rm f}B_{\rm
eq}^2$. Thus, $\alpha_B\approx\lambda/(\eta k_1 k_{\rm f})$. Since
$\lambda$ scales with $u_{\rm rms}k_{\rm f}$ we have $\alpha_B\approx
u_{\rm rms}/(\eta k_1)$, so the quenching coefficient scales with the
magnetic Reynolds number based on the box scale, not the forcing scale
as one might have expected. Taking into account factors of order unity,
e.g.\ $\lambda\approx0.07u_{\rm rms}k_{\rm f}$ (where $k_{\rm f}=5$)
and $(2\pi)^{-1}$, we have $\alpha_B=\eta_B\approx R_{\rm m}/60$, where
$R_{\rm m}=u_{\rm rms}L/\eta$.

We conclude that Eq.~(3) provides an excellent fit to the numerical
simulations (Fig.~1) and one might therefore be tempted to extrapolate
to astrophysical conditions. However, real astrophysical systems have
open boundary conditions and we now need to know whether this could
alleviate the issue of very long timescales for the mean magnetic field.

\end{document}